\documentclass[aps,prl,reprint,superscriptaddress,nobibnotes,nofootinbib,floatfix]{revtex4-2}
\usepackage[english]{babel}
\usepackage{amsmath,amssymb,amsfonts,graphicx,dsfont}
\usepackage[colorlinks,citecolor=blue,urlcolor=blue]{hyperref}
\usepackage{physics}
\usepackage{xcolor}

\newcommand{\kd}{\kappa_{\mathrm{dens}}}

\begin{document}
\renewcommand{\topfraction}{.99}
\renewcommand{\textfraction}{.01}
\renewcommand{\floatpagefraction}{.95}
\setlength{\textfloatsep}{9pt plus 2pt minus 3pt}
\setlength{\dbltextfloatsep}{9pt plus 2pt minus 3pt}

\title{Independent density and coherence skin effects in adaptive fermion circuits}

\author{K. Chahine}
\email{chahine@thp.uni-koeln.de}
\affiliation{Universit\"at zu K\"oln, Institut f\"ur Theoretische Physik,
Z\"ulpicher Str.~77a, 50937 Cologne, Germany}
\author{M. Siegl}
\affiliation{Universit\"at Innsbruck, Institut f\"ur Theoretische Physik,
Technikerstra\ss{}e 21a, 6020 Innsbruck, Austria}
\author{M. Buchhold}
\email{michael.buchhold@uibk.ac.at}
\affiliation{Universit\"at Innsbruck, Institut f\"ur Theoretische Physik,
Technikerstra\ss{}e 21a, 6020 Innsbruck, Austria}

\begin{abstract}
Adaptive circuits use measurements and outcome-conditioned feedback to implement nonreciprocal dynamics without postselection. We show that fermion circuits host independently tunable density and coherence skin effects. The stationary state has an exact density tilt set only by measurement strength and feedback displacement: conditioned-gate details cancel, yielding an exponential one-particle profile and, at any filling, a tilted ensemble of configuration weights. Decaying coherences instead obey, in the bulk, an effective Hatano–Nelson generator controlled by the conditioned gates through a relative occupation phase. Tuning this phase drives a point-gap topological transition and reverses coherence localization at fixed density, allowing density and coherence to localize at opposite edges. We derive both exponents for conditioned \(U(2)\) gates and propose a correlator-ratio probe using ensemble-averaged observables.
\end{abstract}

\maketitle

\emph{Introduction.}---Non-reciprocal hopping can accumulate an extensive
number of eigenstates at a boundary, producing the non-Hermitian skin
effect~\cite{hatano1997,lee2016,alvarez2018,yao2018,kunst2018,leethomale2019,lili2020,zhang2022,ashida2020,bergholtz2021,okumasato2023}. Its origin is
captured by non-Bloch band theory and the point-gap winding
of the bulk spectrum~\cite{yao2018,kunst2018,yokomizo2019,gong2018,kawabata2019,okuma2020,zhang2020,longhi2019,borgnia2020,zirnstein2021}, and has been observed in photonic, electrical, and quantum-walk
platforms~\cite{weidemann2020,helbig2020,hofmann2020,xiaoL2020,ghatak2020,zhangobs2021,liang2022,zhao2025}. When used as a quantum
generator, however, a non-Hermitian Hamiltonian often describes evolution
conditioned on a no-jump record. The physical ensemble instead evolves
under a trace-preserving Liouvillian, which can exhibit skin effects of its
own~\cite{song2019,liu2020,longhi2020,mori2020,haga2021,okumasato2021,nakagawa2021,wanjura2020,yang2022,hamanaka2023}, with a topology that
need not coincide with that of the post-selected
dynamics~\cite{lieu2020,gopalakrishnan2021,kawabata2023,lee2024,chaduteau2026}.

This distinction exposes a question absent at the Hamiltonian level. A
density matrix contains occupations on its diagonal and coherences off of it,
but a non-Hermitian Hamiltonian supplies only one spatial generator for
both. Similarly, canonical asymmetric-hopping Lindbladians provide only a single
control parameter: bond-resolved jumps transport occupations while coherences decay
in place, whereas spatially unresolved jumps translate coherences with the
same bias as the occupations~\cite{endmatter}. Can trace-preserving
dynamics instead endow density and coherence with independently tunable
skin localization, allowing them to select opposite boundaries?

\begin{figure}[t!]
  \centering\includegraphics[width=\linewidth]{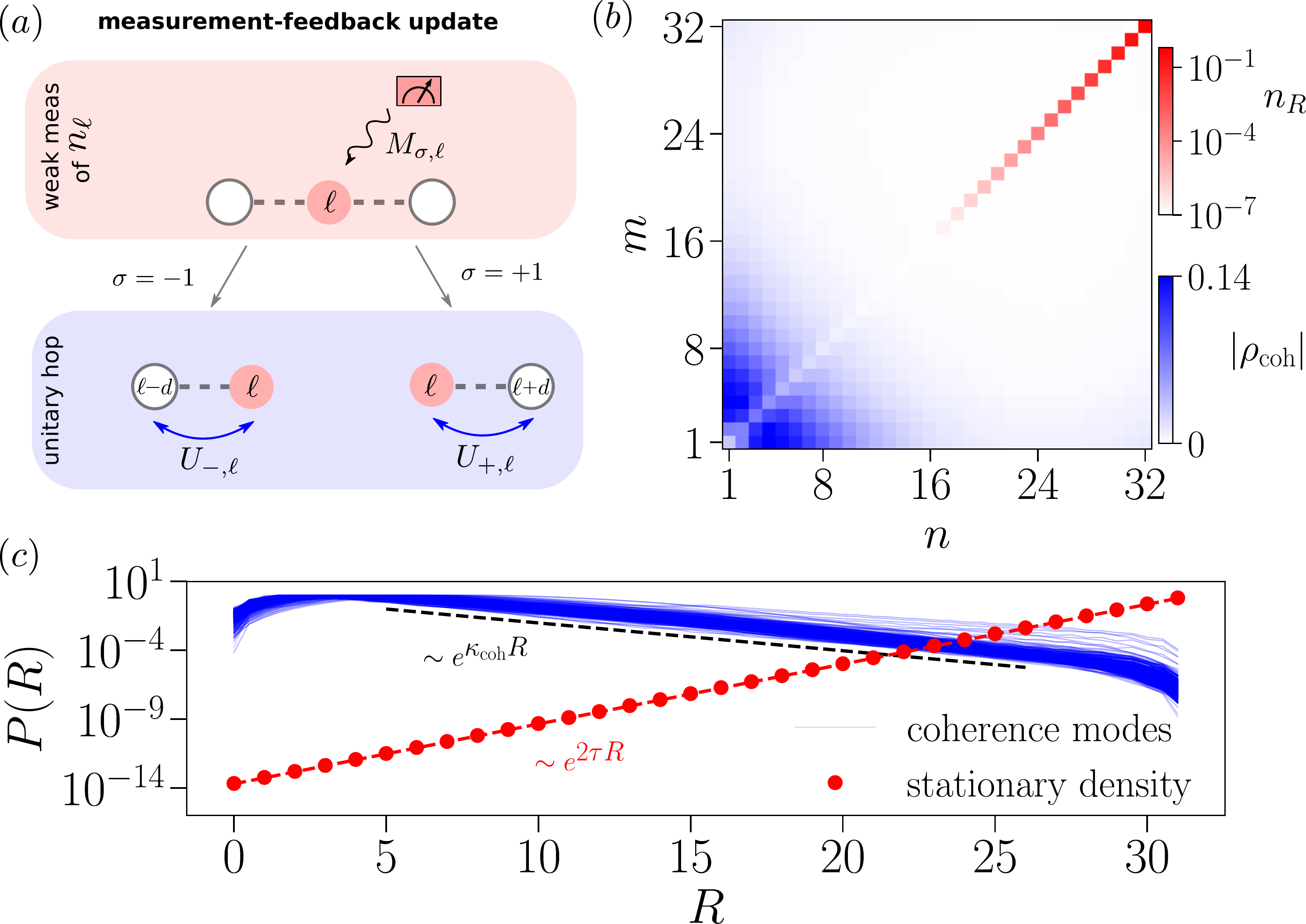}
  \caption{Two skin effects at opposite edges of one adaptive circuit
  ($\tau=0.5$, $L=32$, $d=1$, balanced pair, $\theta=1$). (a) Adaptive measurement-feedback update. A weak measurement of the occupation at site $\ell$ yields $\sigma=\pm1$ and selects a coherent hopping operation on the bond $(\ell-d,\ell)$ for $\sigma=-1$ or $(\ell,\ell+d)$ for $\sigma=+1$.
  (b)~Weight $\langle|\rho_{m,n}|\rangle$ averaged over the coherence modes defined in~\cite{endmatter} (blue) and the stationary state (red), from exact diagonalization of the Liouvillian; each mode is normalized to $\sum_{m,n}|\rho_{m,n}|^2=1$, the stationary state to unit trace, and the two quantities carry separate colour scales.
  (c)~Profiles $P(R)=\sum_{m+n=2R}|\rho_{m,n}|$ of the coherence modes (blue lines) and of the stationary density (red circles); black and red dashed lines: Eqs.~\eqref{eq:kcoh} and~\eqref{eq:xidens}.
}
  \label{fig:fig1}
\end{figure}

We demonstrate that adaptive circuits provide the required dual structure. They combine a
measurement with an outcome-conditioned unitary
gate~\cite{roy2020,sierant2023,iadecola2023,buchhold2022,friedman2023,ravindranath2023,herasymenko2023,piroli2023,odea2024,lemaire2024,ravindranath2025}, an
architecture enabled experimentally by mid-circuit
measurements~\cite{noel2022,hoke2023,koh2023,iqbal2024}. Here we consider an adaptive fermion circuit~\cite{vodenkova2026,steiner2026,sen2026,adachi2022,shenlee2026}, where measurement-feedback implements conditioned hopping processes: measurement amplitudes
set the statistical bias of particle transport, while the phases of
conditioned $U(2)$ gates enter the transfer of coherence [Fig.~\ref{fig:fig1}(a)]. For a large class of unitary feedback gates, these two ingredients produce different non-reciprocities.  

We show that the circuit consequently hosts two distinct
skin effects. The stationary density has the exact exponent
$\kappa_{\rm dens}=2\tau/d$, fixed only by the measurement strength
$\tau$ and feedback displacement $d$ and independent of
$U(2)$ feedback gates. By contrast, an extensive family of decaying
single-particle coherence modes acquires a Hatano--Nelson envelope whose exponent $\kappa_{\rm coh}$ retains
the conditioned-gate phases [Fig.~\ref{fig:fig1}(b)]. Varying a single relative $U(1)$ phase drives
$\kappa_{\rm coh}$ continuously through zero without changing
$\kappa_{\rm dens}$, allowing density and coherence to localize at
opposite boundaries [Fig.~\ref{fig:fig1}(c)]. This reversal is a
point-gap transition of the coherence generator to which the stationary
density is insensitive. We derive both localization exponents and propose
a post-selection-free protocol that extracts $\kappa_{\rm coh}$ from ensemble-averaged correlators.

\emph{Model.}---We consider spinless fermions on a chain of $L$ sites. During an infinitesimal time interval $dt$, a measurement-feedback update is applied independently at each site $\ell$ with probability $\eta dt$. Each update consists of a weak measurement of the local occupation $\hat n_\ell=\hat c_\ell^\dagger\hat c_\ell$ with strength $\tau>0$, described by~\cite{jacobs2006,wiseman2010}
\begin{equation}
\hat M_{\sigma,\ell}
=
\exp\left[\frac{\sigma\tau}{2}(2\hat n_\ell-1)\right]
(2\cosh\tau)^{-\frac12} .
\label{eq:measurement_operator}
\end{equation}
The outcome $\sigma=\pm1$ occurs with Born probability
$p_{\sigma,\ell}=\langle\psi|\hat M_{\sigma,\ell}^2|\psi\rangle$.
The completeness relation
$\sum_\sigma\hat M_{\sigma,\ell}^2=\mathds{1}$
ensures that the probabilities sum to unity. The measurement multiplies an occupied state at $\ell$ by
$\mu_\sigma=e^{\sigma\tau/2}/\sqrt{2\cosh\tau}$
and an empty state by $\mu_{-\sigma}$.

Conditioned on $\sigma$, a unitary gate couples $\ell$ to $\ell+\sigma d$ [Fig.~\ref{fig:fig1}(a)]. We take the gates to be site independent in the bulk. On the ordered pair
$(\beta_0,\beta_1)=(\ell,\ell+\sigma d)$,
$\hat U_{\sigma,\ell}$ is specified by $u^\sigma\in U(2)$ through
\begin{equation}
\hat U_{\sigma,\ell}\hat c^\dagger_{\beta_i}
\hat U^\dagger_{\sigma,\ell}
=\sum_j u^\sigma_{ji}\hat c^\dagger_{\beta_j},
\quad
u^\sigma=e^{i\gamma_\sigma}
\left(\begin{array}{cc}
a_\sigma & b_\sigma\\
-\bar b_\sigma & \bar a_\sigma
\end{array}\right),
\label{eq:gate}
\end{equation}
with all other modes unchanged. Here $(a_\sigma,b_\sigma)$ specifies the $SU(2)$ part of the gate, with
$|a_\sigma|^2+|b_\sigma|^2=1$, and $b_\sigma$ is the transport amplitude. The phase $e^{i\gamma_\sigma}$ in Fock space becomes an occupation phase
$e^{i\gamma_\sigma(\hat n_{\beta_0}+\hat n_{\beta_1})}$. With open boundary conditions (OBC), a gate supported outside of the chain acts as the identity. For $d>1$, the chain decomposes into $d$ independent sublattices.

The measurement and feedback define Kraus operators
$\hat K_{\sigma,\ell}=\hat U_{\sigma,\ell}\hat M_{\sigma,\ell}$, which satisfy
$\sum_\sigma\hat K_{\sigma,\ell}^\dagger\hat K_{\sigma,\ell}=\mathds 1$.
Averaging over outcomes yields the ensemble dynamics
\begin{equation}
\partial_t\hat\rho
=\mathcal{L}\hat\rho=\eta\sum_\ell\left[
\sum_{\sigma=\pm1}
\hat K_{\sigma,\ell}\hat\rho\hat K_{\sigma,\ell}^\dagger
-\hat\rho\right].
\label{eq:qme}
\end{equation}
The associated no-click Hamiltonian is proportional to the identity,
$\hat H_{\rm eff}=-(i\eta/2)\sum_{\ell,\sigma}
\hat K_{\sigma,\ell}^\dagger\hat K_{\sigma,\ell}
=-i\eta L\mathds 1/2$; all nontrivial dynamics therefore resides in the
recycling term. Each $\hat K_{\sigma,\ell}$ conserves particle number, so the
particle-number sectors evolve independently. We focus first on the
one-particle sector and return below to arbitrary filling.

We use two choices of feedback gates. The \emph{conditioned hop} has
$a_\sigma=\cos\varphi$, $b_\sigma=i\sin\varphi$, and
$\gamma_\sigma=0$, yielding
\begin{equation}
\hat U_{\sigma,\ell}
=\exp(i\varphi\hat h_{\ell,\ell+\sigma d}),
\quad
\hat h_{\ell m}=\hat c_\ell^\dagger\hat c_m+\mathrm{h.c.}
\label{eq:conditioned_hop}
\end{equation}
For this choice, both outcomes bias particle transport in the same direction;
$\tau$ controls the nonreciprocity, while $\varphi$ controls the mixing
strength. To tune the coherence skin effect independently, we use the
\emph{balanced pair}
\begin{equation}
a_+=a_-=b_+=\tfrac1{\sqrt2},\quad
b_-=\tfrac{i}{\sqrt2},\quad
\gamma_+-\gamma_-=\theta.
\label{eq:balanced}
\end{equation}
This fixes the mixing probabilities while varying only the relative
occupation phase $\theta$, leaving the stationary density profile unchanged. This allows density and coherences to localize at opposite edges [Fig.~\ref{fig:fig1}(b)].

\emph{Rigid density localization.}---We first determine the stationary one-particle density profile under OBC. Measuring time in units of $\eta^{-1}$, the diagonals $n_R\equiv \rho_{R,R}$ obey
\begin{align}
\partial_t n_R=&
\Gamma_{+}n_{R-d}+\Gamma_{-}n_{R+d}
-(\Gamma_{+}+\Gamma_{-})n_R \nonumber\\
&+2\operatorname{Re}\left[
C\bigl(\rho_{R,R+d}-\rho_{R-d,R}\bigr)\right].
\label{eq:neom}
\end{align}
The first line describes hopping between sites separated by $d$, while the second couples the density to distance-$d$ coherences. The coefficients are
\begin{align*}
\Gamma_{\sigma}&=\mu_\sigma^2\bigl(|b_+|^2+|b_-|^2\bigr),\ C=\mu_+\mu_-\bigl(a_+\bar b_+-\bar a_- b_-\bigr).
\end{align*}

Setting the distance-$d$ coherences in Eq.~\eqref{eq:neom} to zero and taking $n_R\propto\exp(\kappa_{\mathrm{dens}}R)$, the bulk stationarity condition has the roots
$\exp(\kappa_{\mathrm{dens}}d)\in \{1,e^{2\tau}\}$. Under OBC, and provided the gates transport, $|b_+|^2+|b_-|^2>0$, the vanishing current
$J_R=\Gamma_+n_R-\Gamma_-n_{R+d}=0$ selects the tilted root on each residue chain, yielding
\begin{equation}
\frac{n_{R+d}}{n_R}=e^{2\tau},
\quad
\kd=\frac{2\tau}{d},
\quad
\xi_{\mathrm{dens}}\equiv\kd^{-1}=\frac{d}{2\tau}.
\label{eq:xidens}
\end{equation}
It remains to verify that this diagonal ansatz does not source coherences. For the bond $(m,m+d)$, the contributions of the $\sigma=+$ and $\sigma=-$ channels to $\partial_t\rho_{m,m+d}$ are, respectively,
\begin{equation}
-a_+b_+\Delta_m,
\quad
\overline{a_-b_-}\Delta_m,
\quad
\Delta_m=\mu_+^2n_m-\mu_-^2n_{m+d},
\label{eq:source}
\end{equation}
where row orthogonality of $u^\sigma$ was used and its $U(1)$ phase cancels between $\hat K$ and $\hat K^\dagger$. Since
$\mu_+^2/\mu_-^2=e^{2\tau}$, Eq.~\eqref{eq:xidens} gives $\Delta_m=0$. The tilted diagonal profile is therefore an exact stationary state for arbitrary transporting conditioned $U(2)$ gates. Direct diagonalization of the full Liouvillian confirms its exponent [Fig.~\ref{fig:fig1}(c)]. Thus, the stationary density balance is that of a classical biased random walk: the gates set the overall hopping scale, while the bias
$\Gamma_+/\Gamma_-=e^{2\tau}$ is set only by the measurement strength.

The solution extends to arbitrary filling: defining the projector onto $N$ particles $\hat P_{N}$, a stationary state is
\begin{equation}
\hat\rho_{N}=
\frac{1}{Z_{N}}
\hat P_{N}
\exp\left[
\frac{2\tau}{d}\sum_j j\hat n_j
\right].
\label{eq:manybody}
\end{equation}
For any diagonal particle number configurations $\mathcal C$ and $\mathcal C'$ related by an allowed rightward hop of length $d$, their weights obey
$w(\mathcal C')/w(\mathcal C)=e^{2\tau}$.
Thus, the tilt of the configuration weights remains independent of the conditioned gates at arbitrary filling~\cite{endmatter}.

\begin{figure}[t!]  \centering\includegraphics[width=\linewidth]{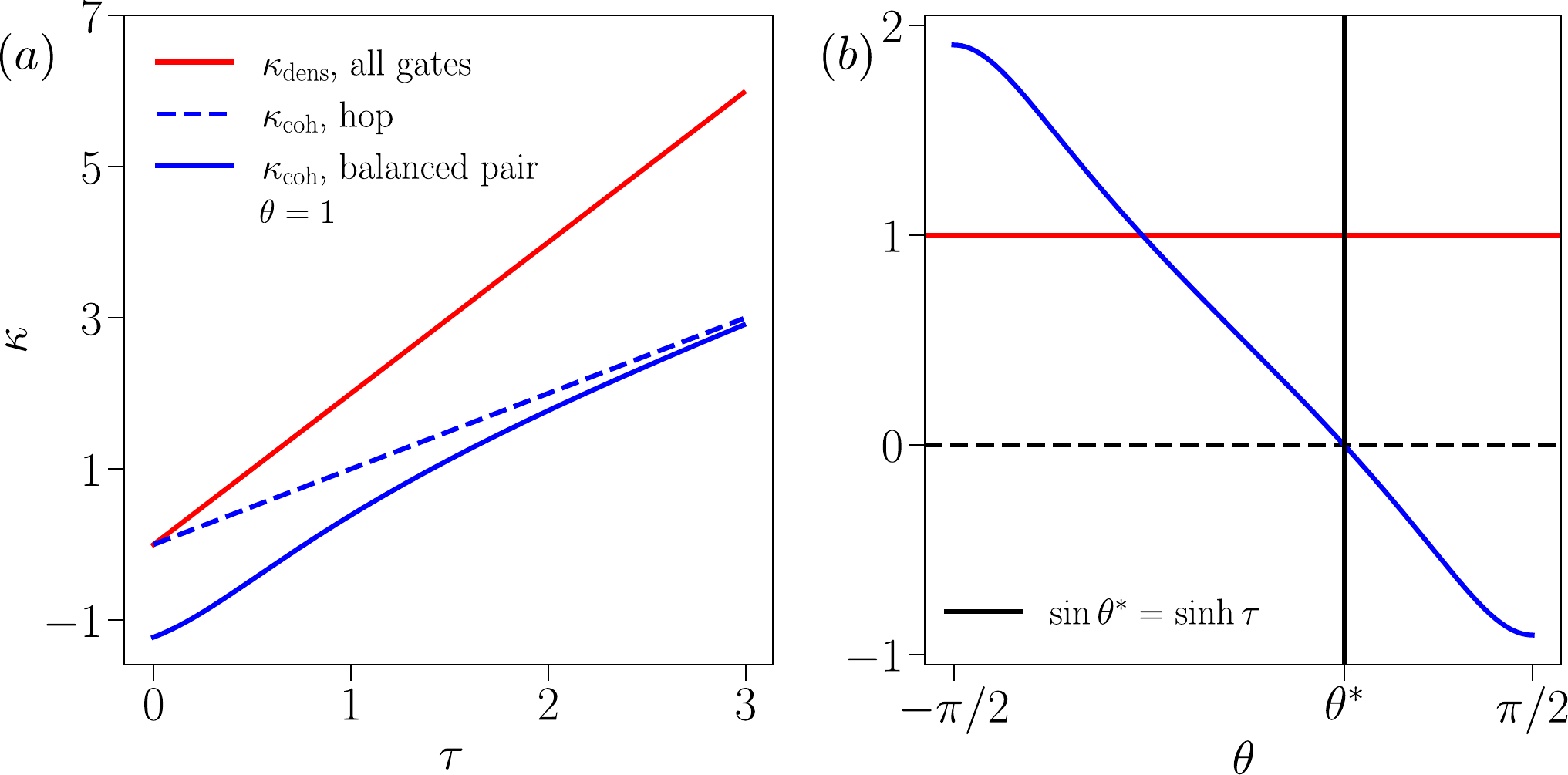}
  \caption{Analytical density (red) and bulk-coherence (blue) exponents against the control parameters of the
  circuit; $d=1$. (a)~Versus measurement
  strength $\tau$. Dashed: conditioned hop, $\kappa_{\rm coh}=\kappa_{\rm dens}/2$. Solid: balanced pair at $\theta=1$. (b)~Versus the phase $\theta$ at
  $\tau=0.5$ for the balanced pair. As
  $\theta=\gamma_+-\gamma_-$ is tuned, $\kappa_\mathrm{coh}$ passes through zero at
  $\sin\theta^{*}=\sinh\tau$, while the density exponent is unaffected.}
  \label{fig:sign}
\end{figure}

\emph{Tunable coherence localization.}---While the stationary state is
diagonal in the occupation-number basis, decaying Liouvillian eigenmodes
carry nonzero coherences. We show that a broad family of these modes also
exhibits a skin effect. Unlike the density skin effect, its localization
length and selected edge can be tuned by the conditioned gates, allowing
the two effects to occur on different length scales and at opposite
boundaries.

Write $\mathcal L=\mathcal L_{\rm bulk}+\mathcal V$, with
$\mathcal L_{\rm bulk}$ chosen to act on the two matrix indices
separately,
\begin{align}
  &(\mathcal L_{\rm bulk}\hat\rho)_{m,n}
  =\bigl(\hat X\hat\rho+\hat\rho\hat X^\dagger\bigr)_{m,n},
  \label{eq:sep}\\
  &\hat X=\frac{a_0}{2}\mathds 1
  +\sum_j\left(t_R\ket{j+d}\bra{j}+t_L\ket{j}\bra{j+d}\right),
  \nonumber
\end{align}
with a uniform onsite term $a_0$~\cite{endmatter} and
\begin{align}
  &t_R=\mu_+B_+,\qquad t_L=\mu_-B_-,\nonumber\\
  &B_\sigma=\mu_\sigma e^{i\gamma_{-\sigma}}b_{-\sigma}
   -\mu_{-\sigma}e^{i\gamma_\sigma}\bar b_\sigma .
  \label{eq:HN_hoppings}
\end{align}
For $m\neq n$ and $|m-n|\neq d$, no gate acts on both indices of $\rho_{m,n}$, and $\mathcal L_{\rm bulk}$ coincides with the full Liouvillian. The remainder $\mathcal V$ contains the contact rows $|m-n|\in\{0,d\}$ and the physical-boundary rows, and is restored below.

Since the two terms of $\mathcal L_{\rm bulk}$ act on different indices,
its eigenmodes are separable,
$\rho^{(\alpha\beta)}_{m,n}=\psi_\alpha(m)\bar\psi_\beta(n)$, with
$\psi_\alpha$, $\psi_\beta$ right eigenvectors of the Hatano--Nelson
matrix $\hat X$~\cite{hatano1997,hatano1997b}. Its hopping asymmetry defines
\begin{equation}
  \kappa_{\rm coh}=\frac1d\ln\left|\frac{t_R}{t_L}\right|
  =\frac1d\left[\tau+\frac12\ln\frac{|B_+|^{2}}{|B_-|^{2}}\right].
  \label{eq:kcoh}
\end{equation}
Let $G=\mathrm{diag}(g^{\ell})$ with $g=(t_R/t_L)^{1/2d}$,
$|g|=e^{\kappa_{\rm coh}/2}$. The two-sided gauge transformation
\begin{equation}
  \hat\rho\mapsto\tilde\rho=G^{-1}\hat\rho\,(G^\dagger)^{-1},
  \qquad
  \rho_{m,n}\mapsto g^{-m}\bar g^{-n}\rho_{m,n},
  \label{eq:gauge}
\end{equation}
maps $\hat X$ to $\tilde X=G^{-1}\hat XG$, a complex symmetric matrix
with equal hoppings $c=\sqrt{t_Rt_L}$ in both directions, and
$\mathcal L_{\rm bulk}$ to $\tilde{\mathcal L}_{\rm bulk}$, which acts as
$\tilde\rho\mapsto\tilde X\tilde\rho+\tilde\rho\tilde X^\dagger$. Both
are normal. The eigenmodes of $\tilde X$ on the open chain are the
hard-wall standing waves $\psi_\alpha(j)\propto\sin k_\alpha j$, with
$|\psi_\alpha(j)|^{2}\le2/(L+1)$ at every site for $d=1$ (and $2/(L_r+1)$ on a residue chain of length $L_r\simeq L/d$), and
$\tilde\rho^{(\alpha\beta)}=\psi_\alpha\psi_\beta^\dagger$ obeys
$\tilde{\mathcal L}_{\rm bulk}\tilde\rho^{(\alpha\beta)}
=(a_0+2c\cos k_\alpha+2\bar c\cos k_\beta)\tilde\rho^{(\alpha\beta)}$.
Every eigenmode of $\mathcal L_{\rm bulk}$ is therefore an extended
standing-wave pattern multiplied by the gauge factor,
\begin{equation}
  |\rho_{m,n}|=e^{\kappa_{\rm coh}R}\,|\tilde\rho_{m,n}|,
  \qquad R=\frac{m+n}{2},
  \label{eq:coherence_envelope}
\end{equation}
so that $\xi_{\rm coh}=1/|\kappa_{\rm coh}|$ is its localization length
along the centre-of-mass coordinate and the sign of $\kappa_{\rm coh}$
selects the edge.

We now restore $\mathcal V$. In the gauged frame
$\tilde{\mathcal L}=\tilde{\mathcal L}_{\rm bulk}+\tilde{\mathcal V}$,
where $\tilde{\mathcal V}$ is supported, in both its rows and columns, on
the defect region formed by the contact set $|m-n|\in\{0,d\}$ and the
physical-boundary rows and columns, and its norm is bounded independently
of $L$~\cite{endmatter}. Let $\hat P$ project onto the defect region and
define, for any $\tilde\rho$, the defect weight
$w_{\rm d}=\|\hat P\tilde\rho\|^2/\|\tilde\rho\|^2$. The gauged problem thus consists of the normal generator
$\tilde{\mathcal L}_{\rm bulk}$ on an $L\times L$ lattice, perturbed on
only $O(L)$ of its $L^2$ sites. If $\tilde{\mathcal L}$ were normal, its
orthonormal eigenbasis would obey
$\sum_\alpha w_{{\rm d},\alpha}=\operatorname{Tr}\hat P=O(L)$.
Hence at most $O(L)$ modes could retain an $O(1)$ defect weight, while
an $O(L^2)$ family would have $w_{\rm d}\to0$. This normal line-defect
result is the analytical origin of the bulk coherence modes. The actual
$\tilde{\mathcal L}$ is non-normal, so this conclusion must be established
separately.

What holds for arbitrary $\tilde{\mathcal V}$ is that its action on an
eigenmode $\tilde{\mathcal L}\tilde\rho=\Lambda\tilde\rho$ is controlled
by $w_{\rm d}$ alone. From
$(\Lambda-\tilde{\mathcal L}_{\rm bulk})\tilde\rho
=\tilde{\mathcal V}\hat P\tilde\rho$,
\begin{equation}
\bigl\|(\Lambda-\tilde{\mathcal L}_{\rm bulk})\tilde\rho\bigr\|
\leq\|\tilde{\mathcal V}\|\sqrt{w_{\rm d}}\,\|\tilde\rho\|,
\label{eq:gaugebounds}
\end{equation}
which uses only the support of $\tilde{\mathcal V}$. For a mode with
$w_{\rm d}\to0$, the residual and, by normality of
$\tilde{\mathcal L}_{\rm bulk}$, the distance of $\Lambda$ from its
bulk spectrum vanish. This bound does not by itself determine the
spatial structure within the corresponding bulk spectral subspace.
Exact diagonalization of the non-normal full Liouvillian supplies this
step: it finds an $O(L^2)$ family whose median defect weight decreases
with $L$ and follows the corresponding bulk-mode value, while its
fitted exponent converges to Eq.~\eqref{eq:kcoh}~\cite{endmatter}.
Modes with finite defect weight need not follow this exponent; the
diagonal stationary state, with $w_{\rm d}=1$, is the extreme case.
Thus, $\kappa_{\rm coh}$ is exact for $\mathcal L_{\rm bulk}$ and
asymptotic for this extensive family of the full Liouvillian. 

For these bulk coherence modes, the gate dependence absent from
$\kappa_{\rm dens}$ survives in $|B_+/B_-|$, with
\begin{equation}
  |B_+|^{2}-|B_-|^{2}=4\mu_+\mu_-\sin(\gamma_+-\gamma_-)\,
  \operatorname{Im}\bigl(\bar b_+\bar b_-\bigr).
  \label{eq:obstruction}
\end{equation}
For gates with a common $U(1)$ phase, including $SU(2)$ gates with
$\gamma_\sigma=0$, the right-hand side vanishes independently of their
$SU(2)$ parameters. It also vanishes whenever
$\operatorname{Im}(\bar b_+\bar b_-)=0$, as for the conditioned hop. In
either case $\kappa_{\rm coh}=\tau/d$ and
$\xi_{\rm dens}=\xi_{\rm coh}/2$. Detuning requires both a relative
$U(1)$ phase and $\operatorname{Im}(\bar b_+\bar b_-)\neq0$. For the
balanced pair at $d=1$, Eq.~\eqref{eq:kcoh} evaluates to
$\kappa_{\rm coh}=\tau+\tfrac12\ln[(\cosh\tau-\sin\theta)/(\cosh\tau+\sin\theta)]$ [Fig.~\ref{fig:sign}(a)].
Varying $\theta$ changes $\kappa_{\rm coh}$ without affecting
$\kappa_{\rm dens}$ and reverses the selected edge at
$\sin\theta^{*}=\sinh\tau$, provided
$\tau\le\operatorname{arsinh}1\simeq0.88$ [Fig.~\ref{fig:sign}(b)].
More generally, opposite-edge localization requires $|B_+/B_-|<e^{-\tau}$.
In the limit $\tau\to0$ the density exponent vanishes, whereas a
phase-carrying balanced pair retains $\kappa_{\rm coh}\neq0$, realizing a coherence skin effect without a
density skin effect.

\emph{Point-gap topology.}---The reversal of coherence localization is a point-gap transition of the Hatano--Nelson matrix $\hat X$ governing either matrix index. Removing the uniform decay, define
$\hat X_0=\hat X-a_0\mathds 1/2$. On each residue chain, its spectrum under periodic boundary conditions (PBC) is
$\lambda_0(\beta)=t_R\beta^{-1}+t_L\beta$, where $\beta=e^{iq}$. This spectrum winds around zero with
\begin{equation}
w=\frac{1}{2\pi i}\oint_{|\beta|=1}d\beta 
\partial_\beta\ln\lambda_0(\beta)
=-\operatorname{sgn}\kappa_{\rm coh}.
\label{eq:wind}
\end{equation}
The two signs of $\kappa_{\rm coh}$ therefore correspond to distinct point-gap phases and cannot be connected without closing the point gap~\cite{gong2018,kawabata2019,okuma2020,zhang2020}. At $\kappa_{\rm coh}=0$, or $|t_R|=|t_L|$, the spectral ellipse collapses onto the OBC segment, the localization length diverges, and the selected edge reverses. For the balanced pair, varying $\theta$ drives this transition at $\sin\theta^*=\sinh\tau$ while leaving $\kappa_{\rm dens}$ unchanged. The winding can persist as $\tau\to0$, where the density skin effect disappears and the dynamics becomes a random-unitary channel.

In contrast to the canonical bond-resolved and unresolved hopping
constructions~\cite{endmatter}, the adaptive circuit
combines two distinct structures: the measurement weights determine the
stationary density tilt, while interference between the conditioned gates
determines the coherence winding. This permits the two sectors to localize
at opposite boundaries.

\begin{figure}[t!]
  \centering\includegraphics[width=\linewidth]{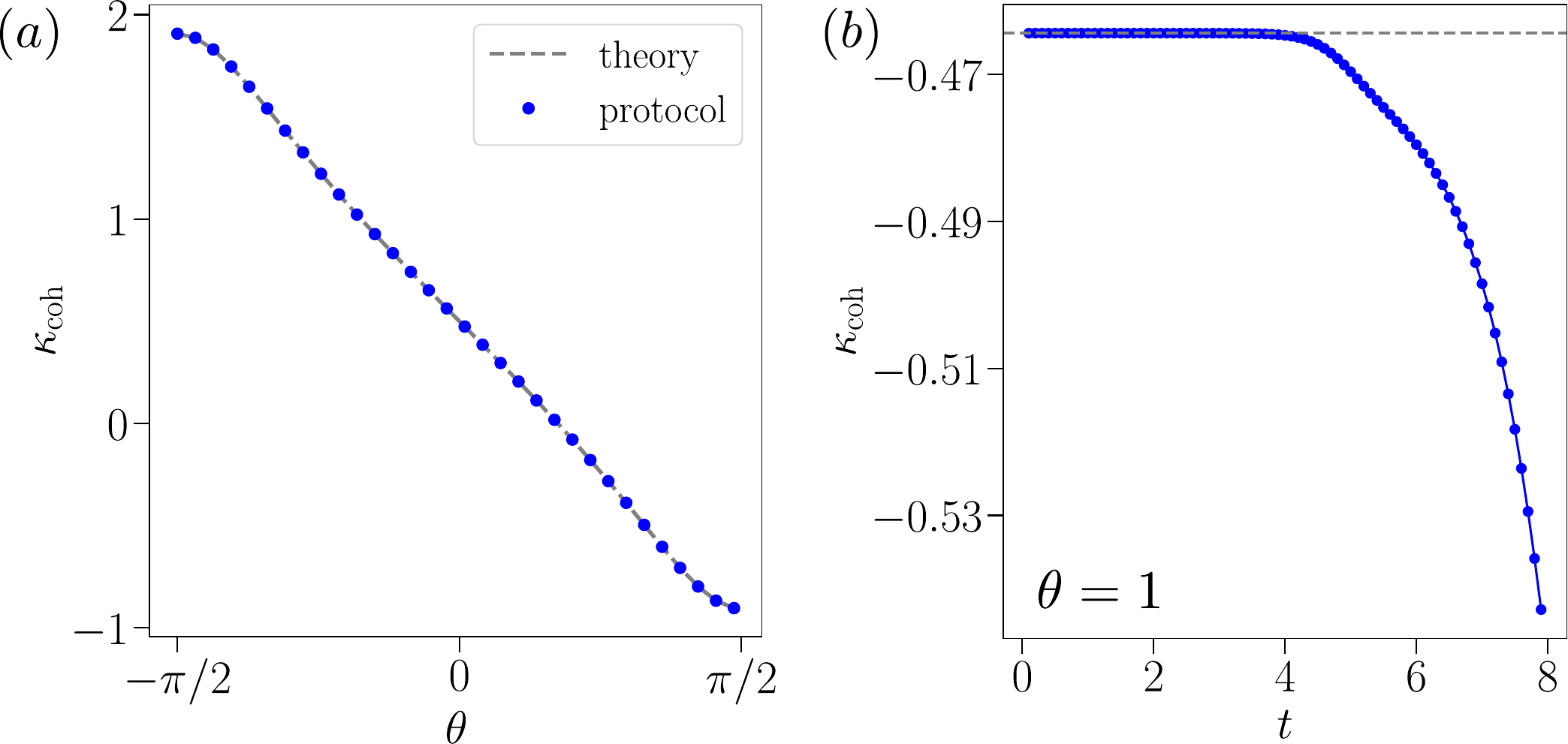}
  \caption{The differential protocol at finite time ($L=28$, $r=6$,
  $\tau=0.5$, $d=1$, balanced pair).
  (a)~Exponent extracted from Eq.~\eqref{eq:ratio} at $t=2$ (circles) and the prediction of Eq.~\eqref{eq:kcoh} (dashed).
  (b)~Extracted exponent against time at $\theta=1$; dashed: Eq.~\eqref{eq:kcoh}.}
  \label{fig:fig4}
\end{figure}

\emph{Detection.}---While $\xi_{\rm dens}$ is read directly from the
stationary occupations $n_R$ in the measured basis, the coherences
vanish identically in that same stationary state. The opposite-edge
localization concerns decaying eigenoperators, which need not be
positive; any physical density matrix obeys
$|\rho_{m,n}|^2\le n_m n_n$. We therefore extract $\xi_{\rm coh}$
from transient, ensemble-averaged correlators without postselection. We prepare two one-particle states, $\ket{\psi_\pm}=(\ket{{m_0}}\pm\ket{{m_0+r}})/\sqrt2$, with $r>d$, and evolve both under the same measurement-feedback circuit. Their difference, $\Delta\rho(t)=\rho_+(t)-\rho_-(t)$, contains only the initially prepared Hermitian coherence, since the diagonal components cancel exactly. We then measure the fixed-separation correlator $f_r(R)=\langle\hat c^\dagger_{R}\hat c_{R+r}\rangle$ for both preparations and form $\Delta f_r(R,t)=f_r^{(+)}(R,t)-f_r^{(-)}(R,t)$. In the bulk, where the coherence pair remains outside the contact set, the evolution is Eq.~\eqref{eq:sep}. The gauge $G^{-1}\hat XG$ of Eq.~\eqref{eq:gauge} maps the non-reciprocal generator to a reciprocal hopping problem; consequently, a localized coherence packet centered at $R_0$ has a symmetric envelope in the transformed basis, while transforming back multiplies its amplitude by $e^{\kappa_{\rm coh}R}$, with $\kappa_{\rm coh}=d^{-1}\ln|t_R/t_L|$. The common time-dependent decay factor likewise cancels in a ratio of symmetric positions, giving
\begin{equation}
\frac{|\Delta f_r(R_0+\Delta R,t)|}
{|\Delta f_r(R_0-\Delta R,t)|}
=e^{2\kappa_{\rm coh}\Delta R}.
\label{eq:ratio}
\end{equation}
Thus two correlators determine $\kappa_{\rm coh}$ without fitting or time calibration. The ingredients, a weak density measurement, feedforward, a conditioned
two-mode gate and one conditional phase, are available in mid-circuit-measurement
architectures~\cite{hoke2023,iqbal2024,shen2026}. Equation~\eqref{eq:ratio} is exact for $\mathcal L_{\rm bulk}$ and accurate for the full Liouvillian in a time window before the packet reaches the contact set or the physical boundaries, where the symmetry of the gauged packet is lost. A simulation of the protocol for the full Liouvillian reproduces Eq.~\eqref{eq:kcoh} at $t=2$ across the whole phase sweep [Fig.~\ref{fig:fig4}(a)]; Fig.~\ref{fig:fig4}(b) shows the extracted exponent leaving the prediction once the packet reaches the boundaries.

\emph{Discussion.}---The adaptive circuit separates two forms of
nonreciprocity. Measurement backaction fixes the exact stationary
density tilt, Eq.~\eqref{eq:xidens}, independently of the conditioned
gates and, as a tilt of configuration weights, at arbitrary filling.
Interference between the conditioned gates instead fixes the bulk
coherence exponent, Eq.~\eqref{eq:kcoh}, and its point-gap winding.
This exponent is exact for $\mathcal L_{\rm bulk}$ and asymptotic for
the extensive family of full-Liouvillian modes whose gauged defect
weight vanishes with system size. 

Two directions follow. First, the coherence exponent derived here is a
one-particle result, because the one-body equations do not close at
finite filling. Nevertheless, individual trajectories underlying
Eq.~\eqref{eq:qme} preserve Gaussianity,
since $\hat M_{\sigma,\ell}$ and $\hat U_{\sigma,\ell}$ are generated
by number-conserving quadratic forms. Ensemble-averaged two-point
functions at finite density can therefore be obtained efficiently by
trajectory sampling~\cite{cao2019,alberton2021}, allowing the
persistence of coherence localization and of the correlator ratio in
Eq.~\eqref{eq:ratio} to be tested beyond the one-particle sector.
Second, as $\tau\to0$, the outcomes become state-independent unbiased
binary variables and the dynamics reduces to a random-unitary circuit
with classical binary control. The coherence skin effect then requires
only two gates with a relative occupation phase selected by this
variable, and can therefore be realized without measurement
backaction.

\begin{acknowledgments}
\emph{Acknowledgements.}-- M.\,S. and M.\,B. acknowledge support from the Austrian Science Fund (FWF)
quantA Cluster of Excellence. M.\,B. acknowledges support from the
Heisenberg programme of the Deutsche Forschungsgemeinschaft (DFG, German
Research Foundation), project no.~549109008. K.\,C. and M.\,B. acknowledge
support from the Deutsche Forschungsgemeinschaft (DFG, German Research
Foundation) under DFG Collaborative Research Center (CRC) 183, Project
No.~277101999, project B01. Code and data reproducing every number in this
Letter are openly available on Zenodo \cite{chahine_2026_22942339}.
\end{acknowledgments}

\section*{End Matter}

In the one-particle sector $\hat M_{\sigma,\ell}|j\rangle=\mu_\sigma|j\rangle$
for $j=\ell$ and $\mu_{-\sigma}|j\rangle$ for $j\neq\ell$, and the gate acts
only on $j\in\{\ell,\ell+\sigma d\}$; for a coherence $\rho_{m,n}$ the
index not in the pair therefore contributes the factor $\mu_{-\sigma}$. Collecting the two channels gives
$\Gamma_\pm$, $C$ and Eq.~\eqref{eq:HN_hoppings}, and the uniform decay of a
distant coherence,
\begin{align}
  a_0=2\,\mathrm{Re}\bigl[&\mu_+\mu_-\bigl(e^{i\gamma_+}a_++e^{i\gamma_-}a_-\bigr)\nonumber\\
  &+\mu_-^{2}e^{i\gamma_+}\bar a_++\mu_+^{2}e^{i\gamma_-}\bar a_-\bigr]-4,
  \label{eq:a0}
\end{align}
real for every gate. For the conditioned hop,
$a_0=2\cos\varphi(1+\mathrm{sech}\,\tau)-4$ and
$t_{R,L}=\tfrac i2\sin\varphi(1+\mathrm{sech}\,\tau\pm\tanh\tau)$, whence
$t_R/t_L=e^{\tau}$. 

\emph{The contact term in the gauged frame.}---The remainder
$\tilde{\mathcal V}=\tilde{\mathcal L}-\tilde{\mathcal L}_{\rm bulk}$ has
three exact properties. (i)~With $\hat P$ the projector onto the defect
region, $\tilde{\mathcal V}=\hat P\tilde{\mathcal V}\hat P$: rows outside
the defect vanish by Eq.~\eqref{eq:sep}, and columns vanish because a
gate reaching outward from the contact band acts on one index only and
therefore enters with exactly the bulk coefficient. (ii)~Each row of
$\tilde{\mathcal V}$ has a fixed number of entries, each a bounded gate
expression times $g^{\Delta m}\bar g^{\Delta n}$ with
$|\Delta m|,|\Delta n|\le d$, so its maximum row and column sums, and
hence $\|\tilde{\mathcal V}\|_2$, are independent of $L$, (iii)~$\tilde{\mathcal L}_{\rm bulk}$ is normal, so
its numerical range is the convex hull $\mathcal P$ of its spectrum
$\{a_0+2c\cos k_\alpha+2\bar c\cos k_\beta\}$, a parallelogram that
collapses to a segment when $t_Rt_L$ is real, as for the conditioned hop.

\begin{figure}[t!]
  \centering\includegraphics[width=\linewidth]{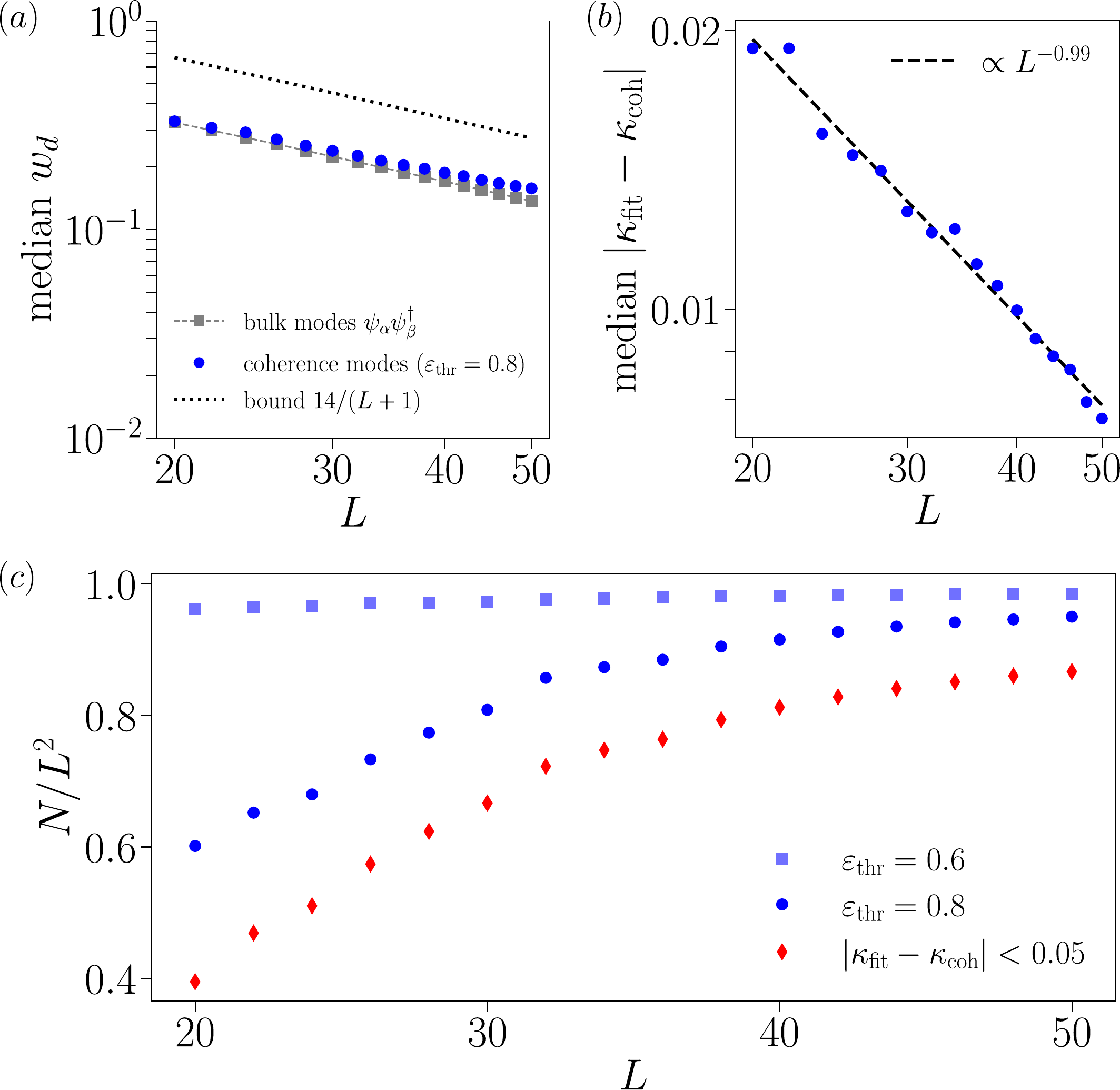}
  \caption{Convergence of the coherence modes of the full open-chain
  Liouvillian to the bulk exponent (balanced pair, $\tau=0.5$,
  $\theta=1$, $d=1$). (a)~Median gauged defect weight $w_{\rm d}$ of the
  coherence modes [Eq.~\eqref{eq:EM_coh_cef}, $\varepsilon_{\rm thr}=0.8$]
  and of the eigenmodes $\psi_\alpha\psi_\beta^\dagger$ of
  $\mathcal L_{\rm bulk}$; dotted: the bound $14/(L+1)$. (b)~Median
  $|\kappa_{\rm fit}-\kappa_{\rm coh}|$ of the coherence modes; dashed:
  power-law fit. (c)~Number $N$ of eigenmodes satisfying
  Eq.~\eqref{eq:EM_coh_cef} for $\varepsilon_{\rm thr}=0.6$ (squares) and
  $0.8$ (circles), and with $|\kappa_{\rm fit}-\kappa_{\rm coh}|<0.05$
  (diamonds), divided by $L^2$.}
  \label{fig:EM_scaling}
\end{figure}

Two bounds follow for an eigenmode
$\tilde{\mathcal L}\tilde\rho=\Lambda\tilde\rho$. Taking the expectation
of $\Lambda=\tilde{\mathcal L}_{\rm bulk}+\tilde{\mathcal V}$ in
$\tilde\rho$ and using (i) and (iii),
$\mathrm{dist}(\Lambda,\mathcal P)\le\|\tilde{\mathcal V}\|\,w_{\rm d}$;
the two-sided support is what makes this linear rather than of order
$\sqrt{w_{\rm d}}$. Applying the spectral theorem to
Eq.~\eqref{eq:gaugebounds},
$\|(1-\hat\Pi_\delta)\tilde\rho\|
\le\|\tilde{\mathcal V}\|\sqrt{w_{\rm d}}\,\|\tilde\rho\|/\delta$, with
$\hat\Pi_\delta$ the projector onto the gauged bulk modes within $\delta$
of $\Lambda$. The first bound has a converse: a mode whose eigenvalue
lies at distance $\Delta$ from $\mathcal P$ has
$w_{\rm d}\ge\Delta/\|\tilde{\mathcal V}\|$. Since $\tilde{\mathcal L}$ is not
normal, neither bound proves that $w_{\rm d}\to0$ for an extensive family,
nor does it fix the fitted exponent of a mode with finite $w_{\rm d}$;
these are the statements Fig.~\ref{fig:EM_scaling} establishes.

For the gauged bulk modes themselves the defect weight is bounded
exactly. With $L_r$ the length of a residue chain ($L_r=L$ for $d=1$),
$|\psi_\alpha(j)|^{2}\le2/(L_r+1)$; a window of $2d+1$ consecutive sites
contains at most three sites of one residue class, so the weight of
$\psi_\alpha\psi_\beta^\dagger$ on the band $|m-n|\le d$, which contains
the contact set, is at most $6/(L_r+1)$, and on the boundary strips of
width $d$ at most $8/(L_r+1)$. Hence $w_{\rm d}\le14/(L_r+1)$ for all
$L^{2}$ of them.

\emph{Numerical scaling of the coherence modes.}---Since
$\tilde{\mathcal L}$ is not normal, the existence of an extensive family
of modes with vanishing defect weight and exponent $\kappa_{\rm coh}$ is
established by exact diagonalization of the full open-chain Liouvillian,
for the balanced pair at the parameters of Fig.~\ref{fig:fig4}, where the
coherence localizes opposite to the density. Right eigenoperators are
normalized to $\sum_{m,n}|\rho_{m,n}|^2=1$, and a mode is counted as a
coherence mode if
\begin{equation}
   \sum_{|m-n|\neq0,d}|\rho_{m,n}|^2>\varepsilon_\mathrm{thr},
   \label{eq:EM_coh_cef}
\end{equation}
i.e.\ if a fraction $\varepsilon_{\rm thr}$ of its weight lies outside the
contact set; $\varepsilon_{\rm thr}=0.8$ in Fig.~\ref{fig:fig1}. For each
mode we compute the gauged defect weight $w_{\rm d}$ and the exponent
$\kappa_{\rm fit}$. The former is computed as $w_{\rm d}=\sum_{m,n\in\mathcal{D}}|\tilde\rho_{m,n}|^2$, where $\mathcal{D}=\{|m-n|=0,d \cup \text{physical boundaries}\}$ and $\tilde\rho$ is an eigenmode in the gauged frame. The latter is extracted from the coherence modes via a linear fit of $\ln P(R)$ with
$P(R)=\sum_{m+n=2R}|\rho_{m,n}|$. The fit is performed over the interior of the
support of $P$ by trimming 20\% of each end. The gauged profile of an eigenmode of
$\mathcal L_{\rm bulk}$ is reflection symmetric, so this fit returns
$\kappa_{\rm coh}$ exactly for the bulk modes  and
$|\kappa_{\rm fit}-\kappa_{\rm coh}|$ measures the asymmetry induced by
the contact term for the eigenmodes of $\mathcal{L}$. Figure~\ref{fig:EM_scaling} shows that the median
defect weight of the coherence modes follows that of the bulk modes,
that their median exponent converges to $\kappa_{\rm coh}$, and that the
number of modes qualifying as coherence modes, as well as the number
whose exponent has converged, approaches $L^2$ with an $O(L)$ deficit.
The modes in this deficit are defect dominated, the stationary state with
$w_{\rm d}=1$ being the extreme case.

\emph{Conventional asymmetric-hopping Lindbladians.}---Consider first the
bond-resolved jumps
\begin{equation}
\hat J_{\sigma,\ell}
=\sqrt{\Gamma_\sigma}\ket{\ell+\sigma}\bra{\ell},
\qquad \sigma=\pm1,
\end{equation}
used in the canonical Liouvillian skin model of
Ref.~\cite{haga2021}. In the bulk they generate
\begin{align}
\partial_t\rho_{m,m}
&=\Gamma_+\rho_{m-1,m-1}+\Gamma_-\rho_{m+1,m+1}
-(\Gamma_++\Gamma_-)\rho_{m,m},\\
\partial_t\rho_{m,n}
&=-(\Gamma_++\Gamma_-)\rho_{m,n},
\qquad m\neq n.
\end{align}
The density therefore performs a biased random walk, while coherences
decay without spatial propagation: for $m\neq n$, no single bond-resolved
jump can act on both sides of $\ket m\bra n$.

At the opposite limit, consider spatially unresolved shift jumps
$\hat J_\sigma=\sqrt{\Gamma_\sigma}\hat S_\sigma$, where
$\hat S_\sigma=\sum_\ell\ket{\ell+\sigma}\bra{\ell}$. Away from the
boundaries,
\begin{equation}
\partial_t\rho_{m,n}
=\Gamma_+\rho_{m-1,n-1}+\Gamma_-\rho_{m+1,n+1}
-(\Gamma_++\Gamma_-)\rho_{m,n}.
\end{equation}
Every diagonal $m-n=\mathrm{const.}$ of the density matrix consequently
undergoes the same biased walk: coherences propagate, but their
localization direction is fixed by the same ratio
$\Gamma_+/\Gamma_-$ as the density. Dressing the shifts with an identity component,
$\hat J_\sigma=\sqrt{\Gamma_\sigma}\,(\alpha\mathds 1+\beta\hat S_\sigma)$,
adds single-index hopping: collecting the terms linear in $\hat S_\sigma$
from $\hat J_\sigma\hat\rho\hat J_\sigma^\dagger$ and from
$-\tfrac12\{\hat J_\sigma^\dagger\hat J_\sigma,\hat\rho\}$, and using
$\hat S_\sigma^\dagger=\hat S_{-\sigma}$ in the bulk, the left index
hops to the right with amplitude
$t_R=\tfrac12(\bar\alpha\beta\,\Gamma_+-\alpha\bar\beta\,\Gamma_-)$ and to
the left with $t_L=\tfrac12(\bar\alpha\beta\,\Gamma_--\alpha\bar\beta\,\Gamma_+)=-\bar t_R$.
Hence $|t_R|=|t_L|$ for any $\alpha,\beta$, and the dressing does not
provide an independent coherence bias.

\emph{Stationary state at arbitrary filling.}---
We derive Eq.~\eqref{eq:manybody} by first constructing
a stationary product state. Consider
\begin{equation}
  \hat\rho_{\boldsymbol\lambda}
  =
  \frac{1}{\mathcal Z_{\boldsymbol\lambda}}
  \prod_{j=1}^{L}e^{\lambda_j\hat n_j},
  \qquad
  \mathcal Z_{\boldsymbol\lambda}
  =
  \prod_{j=1}^{L}(1+e^{\lambda_j}),
  \label{eq:em_product}
\end{equation}
with real coefficients $\lambda_j$ to be determined.
The ensemble-averaged dynamics is
\begin{equation}
  \partial_t\hat\rho
  =
  \eta\sum_\ell
  \left[
    \sum_{\sigma=\pm1}
    \hat K_{\sigma,\ell}\hat\rho
    \hat K_{\sigma,\ell}^\dagger
    -\hat\rho
  \right].
  \label{eq:em_qme}
\end{equation}

Fix an outcome $\sigma$ and the ordered pair
$(\ell,\ell+\sigma d)$.
We denote the unnormalized density operator immediately
after the measurement by
\begin{equation}
  \hat D_{\sigma,\ell}
  \equiv
  \hat M_{\sigma,\ell}
  \hat\rho_{\boldsymbol\lambda}
  \hat M_{\sigma,\ell}.
  \label{eq:em_measured_state}
\end{equation}
For a fixed occupation pattern $\nu$ outside the pair,
let $D_{\sigma,\ell;\nu}^{(1)}$ be its matrix restricted
to the ordered basis
$(|\nu;10\rangle,|\nu;01\rangle)$.
The first local occupation refers to the measured site
$\ell$. Using the measurement amplitudes $\mu_\sigma$
defined in the main text, we obtain
\begin{equation}
  D_{\sigma,\ell;\nu}^{(1)}
  =
  C_\nu
  \begin{pmatrix}
    \mu_\sigma^2e^{\lambda_\ell} & 0\\
    0 & \mu_{-\sigma}^2e^{\lambda_{\ell+\sigma d}}
  \end{pmatrix},
  \label{eq:em_measured_block}
\end{equation}
where
\begin{equation}
  C_\nu
  =
  \frac{1}{\mathcal Z_{\boldsymbol\lambda}}
  \exp\!\left[
    \sum_{j\notin\{\ell,\ell+\sigma d\}}
    \lambda_j n_j^\nu
  \right]
\end{equation}
is the common exterior weight.

The feedback conjugates this block by $u^\sigma$.
It leaves the block invariant whenever its diagonal
entries coincide, since the block is then proportional
to the identity. This gives
\begin{equation}
  \mu_\sigma^2e^{\lambda_\ell}
  =
  \mu_{-\sigma}^2e^{\lambda_{\ell+\sigma d}},
  \qquad
  \lambda_{\ell+\sigma d}-\lambda_\ell
  =
  2\sigma\tau.
  \label{eq:em_balance}
\end{equation}
On each open residue chain, the solution is
\begin{equation}
  \lambda_j
  =
  \frac{2\tau}{d}j+c_{r(j)},
  \qquad
  r(j)=j\bmod d,
  \label{eq:em_tilt}
\end{equation}
where $c_r$ is an arbitrary constant for each residue
class.

The empty and doubly occupied local sectors are
one-dimensional. Their gate eigenvalues are $1$ and
$\det u^\sigma$, respectively, and these phases cancel
in the density operator. Consequently, for the
coefficients in Eq.~\eqref{eq:em_tilt},
\begin{equation}
  \hat U_{\sigma,\ell}
  \hat D_{\sigma,\ell}
  \hat U_{\sigma,\ell}^\dagger
  =
  \hat D_{\sigma,\ell}.
\end{equation}
For an out-of-range partner, the gate acts as the
identity, so the same relation holds at the boundaries.
Since the product state commutes with the measurement
operators, completeness gives
\begin{align}
  \sum_{\sigma=\pm1}
  \hat K_{\sigma,\ell}\hat\rho_{\boldsymbol\lambda}
  \hat K_{\sigma,\ell}^\dagger
  &=
  \sum_{\sigma=\pm1}\hat D_{\sigma,\ell}
  \nonumber\\
  &=
  \hat\rho_{\boldsymbol\lambda}
  \sum_{\sigma=\pm1}\hat M_{\sigma,\ell}^2
  =
  \hat\rho_{\boldsymbol\lambda}.
\end{align}
Thus every site contribution to
Eq.~\eqref{eq:em_qme} vanishes separately.

To restrict the state to fixed particle numbers,
define
\begin{equation}
  \hat N_r=\sum_{j:\,r(j)=r}\hat n_j
\end{equation}
and let $P_{\mathbf N}$ project onto the sector with
$\hat N_r=N_r$ for every residue class.
All Kraus operators commute with $P_{\mathbf N}$,
so the normalized projection of a stationary state
is stationary:
\begin{equation}
  \hat\rho_{*,\mathbf N}
  =
  \frac{
    P_{\mathbf N}\hat\rho_{\boldsymbol\lambda}P_{\mathbf N}
  }{
    \operatorname{Tr}
    (P_{\mathbf N}\hat\rho_{\boldsymbol\lambda})
  }.
\end{equation}
Within this sector, the factor
$\exp(\sum_r c_r\hat N_r)$ is a scalar and cancels
upon normalization. Choosing $c_r=0$ and summing over all
$\mathbf N$ with $\sum_rN_r=N$, i.e.\ projecting with the
total-number projector $\hat P_N=\sum_{\mathbf N}P_{\mathbf N}$, yields
Eq.~\eqref{eq:manybody},
with
\begin{equation}
  Z_{N}
  =
  \operatorname{Tr}\!\left[
    P_{N}
    \exp\!\left(
      \frac{2\tau}{d}\sum_j j\,\hat n_j
    \right)
  \right].
\end{equation}

The unprojected product state has exact Fermi occupations,
\begin{equation}
  n_j
  =
  \frac{e^{\lambda_j}}{1+e^{\lambda_j}},
  \qquad
  \frac{n_j}{1-n_j}
  =
  e^{c_{r(j)}}e^{2\tau j/d}.
  \label{eq:em_fermi}
\end{equation}
Here $e^{c_r}$ is the fugacity of residue chain $r$. For $d>1$ the
$\hat N_r$ are separately conserved, so Eq.~\eqref{eq:manybody} is
stationary but not the unique stationary state of the total-$N$ sector.
Projection generally changes the one-body density,
so Eq.~\eqref{eq:em_fermi} is exact for the unprojected
ensemble, but not generally for a finite system at
fixed particle number.

\bibliography{prl_refs}

\end{document}